\documentclass[twocolumn,showpacs,superscriptaddress,amssymb,10pt,pra,floatfix,aps,longbibliography]{revtex4-1}

\usepackage{graphicx}% Include figure files
\usepackage{amsmath}
\usepackage{dcolumn}% Align table columns on decimal point
\usepackage{bm}% bold math
\usepackage{epsfig}
\usepackage{color}
\usepackage{longtable}
\usepackage{leftidx,amsmath}
\usepackage{natbib} 
\usepackage[utf8]{inputenc}
\usepackage{braket}
\usepackage{tikz}
\usetikzlibrary{snakes}
\usepackage{makecell}
\usepackage[flushleft]{threeparttable}
\usepackage[utf8]{inputenc}
\usetikzlibrary{snakes}
\allowdisplaybreaks

\def\mem#1#2#3{  \left\langle #1 \left\vert  #2 \right\vert #3 \right\rangle   }

\def\emc2{\ensuremath{~m_\text{e}c^2}}

\begin{document}
	\preprint{}
	%
%
%-----------------------------------------Title---------------------------------------
%
%
	\title{Calculations of Delbrück scattering to all orders in $\alpha Z$}
%
%
%-----------------------------------------Author--------------------------------------
%
%

	\author{J.~Sommerfeldt}
	\email{j.sommerfeldt@tu-braunschweig.de}
	\affiliation{Physikalisch--Technische Bundesanstalt, D--38116 Braunschweig, Germany}
	\affiliation{Technische Universit\"at Braunschweig, D--38106 Braunschweig, Germany}
	
	\author{V.~A.~Yerokhin}
    \affiliation{Center for Advanced Studies, Peter the Great St. Petersburg State Polytechnic University, 195251 St.  Petersburg, Russia}
	
	\author{R.~A.~M{\"u}ller}
	\affiliation{Physikalisch--Technische Bundesanstalt, D--38116 Braunschweig, Germany}
	\affiliation{Technische Universit\"at Braunschweig, D--38106 Braunschweig, Germany}
	
	\author{V.~A.~Zaytsev}
	\affiliation{Department of Physics, St.~Petersburg State University, 199034 St.~Petersburg, Russia}

	\author{A.~V.~Volotka}
	\affiliation{School of Physics and Engineering, ITMO University, 197101 St.
Petersburg, Russia}
	
	\author{A.~Surzhykov}
	\affiliation{Physikalisch--Technische Bundesanstalt, D--38116 Braunschweig, Germany}
	\affiliation{Technische Universit\"at Braunschweig, D--38106 Braunschweig, Germany}	

	\date{\today \\[0.3cm]}

%
%
%-----------------------------------------Abstract--------------------------------------
\begin{abstract}
We present a theoretical method to calculate Delbrück scattering amplitudes. Our formalism is based on the exact analytical Dirac-Coulomb Green's function and, therefore, accounts for the interaction of the virtual electron-positron pair with the nucleus to all orders, including the Coulomb corrections. The numerical convergence of our calculations is accelerated by solving the radial integrals that are involved analytically in the asymptotic region. Numerical results for the collision of photons with energies 102.2 keV and 255.5 keV with bare neon and lead nuclei are compared with the predictions of the lowest-order Born approximation. We find that our method can produce accurate results within a reasonable computation time and that the Coulomb corrections enhance the absolute value of the Delbrück amplitude by a few percent for the studied photon energies.
\end{abstract}
\maketitle

\section{Introduction}
Delbrück scattering is the process in which a photon is elastically scattered by the Coulomb field of an atomic nucleus via the formation of virtual electron-positron pairs \cite{meitner_uber_1933}. It is one of the few non-linear quantum electrodynamical processes that is observed experimentally \cite{PhysRevD.8.3813, PhysRevC.23.1375, Muckenheim_1980}. Moreover, precise knowledge of Delbrück amplitudes is needed in order to extract relevant information from nuclear Compton scattering experiments \cite{MILSTEIN1994183, Arenhovel1986}. Hence, this process has attracted considerable attention both from the experimental and theoretical side in the past.  However, despite the strong motivation for the theoretical analysis of Delbrück scattering, most studies have been limited to approximations of the coupling between the virtual electron-positron pair and the nucleus. For example, many studies have been conducted using the lowest-order Born approximation, which neglects terms of the order $(\alpha Z)^4$ and higher in the amplitude \cite{PhysRevD.12.206, PhysRevD.26.908, BARNOY1977132, FALKENBERG19921, PhysRevLett.118.204801}. Depending on the energy, this approximation may break down for higher nuclear charges which are of particular experimental interest.  There have also been some efforts to calculate high energy, large and small angle approximations \cite{PhysRev.182.1873, PhysRevD.2.2444, PhysRevD.5.3077, Milstein_1988} and, more recently, the Coulomb corrections in the low energy limit have been calculated by Kirilin and Terekhov \cite{PhysRevA.77.032118}. To the best of our knowledge, exact calculations (in $\alpha Z$) of Delbrück scattering have only been attempted once  almost three decades ago by Scherdin and co-workers~\cite{PhysRevD.45.2982, scherdin_coulomb_1995}. Due to computational limitations, these calculations had limited numerical accuracy and applicability.

In this work, we present a new method to calculate Delbrück scattering amplitudes that is exact in $\alpha Z$. We employ relativistic quantum electrodynamics, whose basic equations are recalled in Sec.~\ref{Theory}. In Sec.~\ref{EvInt}, we discuss how we evaluate the multidimensional integrals that are involved in the calculation. In particular, in Sec.~\ref{renorm}, we explain how to obtain a finite amplitude by subtracting the free-loop contribution. Moreover, in Sec.~\ref{EInt}, we discuss the analytical structure of the integrand and the rotation of the integration contour. In Sec.~\ref{radInt}, we show how to accelerate the radial integration by solving the integral analytically in the asymptotic case. In Sec.~\ref{compDet}, we discuss the implementation of our method and the magnitude of all possible numerical errors. Numerical results for the collision of photons with energies 102.2 keV and 255.5 keV with bare neon and lead nuclei are compared with the predictions of the lowest-order Born approximation in Sec.~\ref{numResults}. Finally, in Sec. \ref{sumAOut}, we summarize our work and give an outlook to possible applications of our method in the future. Relativistic units $\hbar = m_e = c = 1$ are used throughout this paper, if not stated otherwise.

\section{QED description of Delbrück scattering} \label{Theory}
\subsection{General formalism} \label{QEDdescr}

\begin{figure} [hb]
\begin{center}
\includegraphics[width=0.85\linewidth]{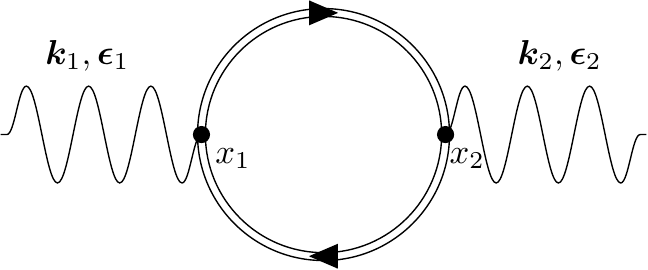}\caption{Leading-order Feynman diagram for Delbrück scattering.}\label{FeynmanDel}
\end{center}
\end{figure}

In the framework of quantum electrodynamics, Delbrück scattering can be described by the Feynman diagram in Fig. \ref{FeynmanDel}. As usual, the wavy lines represent the incoming and outgoing photons with wave vectors  $\boldsymbol{k}_1$ and $\boldsymbol{k}_2$ as well as polarization vectors $\boldsymbol{\epsilon}_1$ and $\boldsymbol{\epsilon}_2$, respectively. Moreover, the double lines describe the virtual electron-positron pair in the Coulomb field of a nucleus. By making use of the well-known Feynman correspondence rules, we can write the amplitude for this diagram as

\begin{equation} \label{MatrixElement}
\begin{aligned}
M &= \frac{i\alpha}{2\pi} \int_{-\infty}^\infty \text{d}z~\int_{-\infty}^\infty \text{d}z'~\int \text{d}^3\boldsymbol{r}_1~\int \text{d}^3\boldsymbol{r}_2~\\
&\times\text{Tr}\Big[\hat{R}(\boldsymbol{r}_1,\boldsymbol{k}_1,\boldsymbol{\epsilon}_1) G(\boldsymbol{r}_1,\boldsymbol{r}_2,z)\hat{R}^\dagger(\boldsymbol{r}_2,\boldsymbol{k}_2,\boldsymbol{\epsilon}_2)\\
&\times G(\boldsymbol{r}_2,\boldsymbol{r}_1,z')\Big]\delta (\omega+z-z')~,\\
\end{aligned}
\end{equation}

\noindent where $z$ and $z'$ are the energy arguments of the electron propagators, $\boldsymbol{r}_1$ and $\boldsymbol{r}_2$ are three-dimensional coordinate vectors and $\omega$ is the energy of the incoming/outgoing photon, see~\cite{MILSTEIN1994183}. Moreover, in Eq. \eqref{MatrixElement}, $\hat{R}(\boldsymbol{r},\boldsymbol{k},\boldsymbol{\epsilon})$ is the photon-lepton interaction operator and $G(\boldsymbol{r}_2,\boldsymbol{r}_1,z')$ is the Dirac-Coulomb Green's function. Both, $\hat{R}(\boldsymbol{r},\boldsymbol{k},\boldsymbol{\epsilon})$ and $G(\boldsymbol{r}_2,\boldsymbol{r}_1,z')$, have to be specified to evaluate the Delbrück amplitude \eqref{MatrixElement} further.

\subsection{Photon-lepton interaction operator}

\noindent The operators $\hat{R}(\boldsymbol{r}_1,\boldsymbol{k_1},\boldsymbol{\epsilon}_1)$ and $\hat{R}^\dagger(\boldsymbol{r}_2,\boldsymbol{k_2},\boldsymbol{\epsilon}_2)$ describe the absorption and emission of photons with wave vectors  $\boldsymbol{k}_1$,  $\boldsymbol{k}_2$ and polarization vectors $\boldsymbol{\epsilon}_1$, $\boldsymbol{\epsilon}_2$. As usual in atomic structure calculations, it is convenient to expand these operators in their multipole components, which have well-defined parity and symmetry properties. By choosing Coulomb gauge for the lepton-matter coupling and by describing the photon polarization in the helicity basis $\boldsymbol{\epsilon} = \boldsymbol{\epsilon}_\lambda = \boldsymbol{e}_x + i \lambda \boldsymbol{e}_y,~\lambda = \pm 1$, this expansion can be written in the form

\begin{equation} \label{ElPhOp}
\begin{aligned}
\hat{R}(\boldsymbol{r},\boldsymbol{k},\boldsymbol{\epsilon}) = \boldsymbol{\alpha}\cdot\boldsymbol{\epsilon}_\lambda e^{i\boldsymbol{k}\cdot\boldsymbol{r}}= &\sqrt{2\pi}\sum_{PLM} i^L  \sqrt{2L+1}\\
&\times (i\lambda)^P D^L_{M\lambda}(\boldsymbol{\hat{k}}) \boldsymbol{\alpha}\cdot \boldsymbol{a}_{LM}^{(P)}~.
\end{aligned}
\end{equation}

\noindent Here, the magnetic ($P = 0$) and electric ($P = 1$) multipole fields read as

\begin{subequations} \label{MultComp}
\begin{align} 
\boldsymbol{a}^{(0)}_{LM} &=  j_{L}(\omega r)~\boldsymbol{T}_{LLM}~,\\
\begin{split}
\boldsymbol{a}^{(1)}_{LM} &= \sqrt{\frac{L+1}{2L+1}}j_{L-1}(\omega r)~\boldsymbol{T}_{L,L-1,M}\\
& - \sqrt{\frac{L}{2L+1}}j_{L+1}(\omega r)~\boldsymbol{T}_{L,L+1,M}~,
\end{split}
\end{align}
\end{subequations}

\noindent where $j_L$ is the spherical Bessel function, and the vector spherical harmonics $T_{JLM}$ are constructed as irreducible tensors of rank $J$ as

\nopagebreak
\begin{equation}
\boldsymbol{T}_{JLM} = \sum_{\mu = -1}^1 \langle L~(M-\mu)~1~\mu\vert J~M\rangle Y_{L,M-\mu}\boldsymbol{\xi}_\mu~.
\end{equation}

\noindent Further details about the multipole representation of the photon-lepton interaction operator can be found in Ref.~\cite{rose_elementary_1957}.

\subsection{Green's function}

Besides the photon-lepton interaction operator, we also have to find a suitable representation of the Green's function $G(\boldsymbol{r}_2,\boldsymbol{r}_1,z)$. Within the well established spectral representation, this function can be written in terms of the eigensolutions of the Dirac equation

\begin{equation} \label{Spect}
G(\boldsymbol{r}_2,\boldsymbol{r}_1,z) = \sum_{n\kappa\mu} \frac{\phi_{n\kappa}^\mu(\boldsymbol{r}_2)\phi_{n\kappa}^{\mu\dagger}(\boldsymbol{r}_1)}{E_{n\kappa}-z}~.
\end{equation}

\noindent Here, $\phi_{n\kappa}^\mu$ are the well-known Dirac-Coulomb wave functions which are conveniently written in the bispinor form

\begin{equation} \label{DCWaveFct}
\phi_{n\kappa}^\mu(\boldsymbol{r}) = \left(\begin{array}{c}
g_{n\kappa}(r) \chi_\kappa^\mu(\boldsymbol{\hat{r}})\\
i f_{n\kappa}(r) \chi_{-\kappa}^\mu(\boldsymbol{\hat{r}})
\end{array}\right)~,
\end{equation}

\noindent with $\kappa$ being the Dirac angular-momentum quantum number, $\mu$ is the projection of the total angular momentum $j = |\kappa|-\frac{1}{2}$ and $n$ is the principal quantum number. Furthermore, $g_{n\kappa}(r)$ and $f_{n\kappa}(r)$ are the large and small radial components and $\chi_\kappa^\mu$ are the spin-angular wave functions \cite{eichler_lectures_2005}. By inserting Eq. \eqref{DCWaveFct} into the spectral representation~\eqref{Spect}, we obtain the Green's function in the form

\begin{widetext}
\begin{equation} \label{GreenSpec}
\begin{aligned}
G(\boldsymbol{r}_2,\boldsymbol{r}_1,z) &= \sum_{n\kappa\mu}\frac{1}{{E_{n\kappa}-z}}\left(\begin{array}{cc}
g_{n\kappa}(r_2)\chi_\kappa^\mu (\boldsymbol{\hat{r}}_2) g_{n\kappa}(r_1)\chi_\kappa^{\mu\dagger} (\boldsymbol{\hat{r}}_1) & -ig_{n\kappa}(r_2)\chi_\kappa^\mu (\boldsymbol{\hat{r}}_2) f_{n\kappa}(r_1)\chi_{-\kappa}^{\mu\dagger} (\boldsymbol{\hat{r}}_1)\\
if_{n\kappa}(r_2)\chi_{-\kappa}^\mu (\boldsymbol{\hat{r}}_2) g_{n\kappa}(r_1)\chi_\kappa^{\mu\dagger} (\boldsymbol{\hat{r}}_1) & f_{n\kappa}(r_2)\chi_{-\kappa}^\mu (\boldsymbol{\hat{r}}_2) f_{n\kappa}(r_1)\chi_{-\kappa}^{\mu\dagger} (\boldsymbol{\hat{r}}_1)
\end{array}\right)~.
\end{aligned}
\end{equation}
\end{widetext}

\noindent As usual, the sum over $n$ is understood here as a summation over the bound discrete states and an integration over the positive and negative continua. 

In Eq.~\eqref{GreenSpec}, the summation (integration) over $n$ runs over the complete Dirac spectrum. This summation is a rather complicated task which needs to be discussed in detail. To start with this discussion, we introduce the radial Green's function

\begin{equation}
G_\kappa(r_2,r_1,z) = \left(\begin{array}{cc}
\sum_{n}\frac{g_{n\kappa}(r_2)g_{n\kappa}(r_1)}{{E_{n\kappa}-z}}&  \sum_{n}\frac{g_{n\kappa}(r_2)f_{n\kappa}(r_1)}{{E_{n\kappa}-z}}\\
\sum_{n}\frac{f_{n\kappa}(r_2)g_{n\kappa}(r_1)}{{E_{n\kappa}-z}}& \sum_{n}\frac{f_{n\kappa}(r_2)f_{n\kappa}(r_1)}{{E_{n\kappa}-z}}
\end{array}\right).
\end{equation}

\noindent To find the components of this function, we can use the fact that $G_\kappa(r_2,r_1,z)$ is the solution of the inhomogeneous equation

\begin{equation}
\begin{aligned}
\left(\begin{array}{cc}
1 + V(r_2) - z & -\frac{1}{r_2} \frac{\text{d}}{\text{d}r_2} r_2 + \frac{\kappa}{r_2}\\
\frac{1}{r_2} \frac{\text{d}}{\text{d}r_2} r_2 + \frac{\kappa}{r_2} & -1 + V(r_2) - z
\end{array}\right) G_\kappa(r_2,r_1,z)\\
 = \left(\begin{array}{cc}
1\hphantom{aa} & 0\\
0\hphantom{aa} & 1
\end{array}\right)\frac{\delta (r_2-r_1)}{r_2r_1}~,
\end{aligned}
\end{equation}

\noindent where in the second line $\delta(x)$ is the Dirac delta function \cite{hylton_reduced_1984, MOHR1998227}. We can find the solutions of this equation as

\begin{equation}
\begin{aligned}
G_\kappa(&r_2,r_1,z) = \frac{1}{w_\kappa(z)} \\
&\times\Bigg[\Theta(r_2-r_1) \left(\begin{array}{c}
F_{\kappa,\infty}^1 (r_2,z)\\
F_{\kappa,\infty}^2 (r_2,z)\\
\end{array}\right) \left(\begin{array}{c}
F_{\kappa,0}^1 (r_1,z)\\
F_{\kappa,0}^2 (r_1,z)\\
\end{array}\right)^T\\
&+\Theta(r_1-r_2) \left(\begin{array}{c}
F_{\kappa,0}^1 (r_2,z)\\
F_{\kappa,0}^2 (r_2,z)\\
\end{array}\right) \left(\begin{array}{c}
F_{\kappa,\infty}^1 (r_1,z)\\
F_{\kappa,\infty}^2 (r_1,z)\\
\end{array}\right)^T\Bigg]~,
\end{aligned}
\end{equation}

\noindent where the index $T$ indicates the transpose and the Wronskian

\begin{equation} \label{Wronskian}
w_\kappa(z) = r^2 [F^2_{\kappa,0}(r,z) F^1_{\kappa,\infty}(r,z)-F^1_{\kappa,0}(r,z) F^2_{\kappa,\infty}(r,z)]~,
\end{equation} 

\noindent is known to be independent of $r$. Moreover, the functions $F^{1,2}_{\kappa,0}(r)$ and $F^{1,2}_{\kappa,\infty}(r)$ are the solutions of the homogeneous equation

\begin{equation}
\left(\begin{array}{cc}
1 + V(r) - z & -\frac{1}{r} \frac{\text{d}}{\text{d}r} r + \frac{\kappa}{r}\\
\frac{1}{r} \frac{\text{d}}{\text{d}r} r + \frac{\kappa}{r} & -1 + V(r) - z
\end{array}\right)\left(\begin{array}{c}
F_{\kappa}^1 (r,z)\\
F_{\kappa}^2 (r,z)\\
\end{array}\right)  = 0 ~,
\end{equation} 

\noindent that are regular at the origin and at infinity, see Refs.~\cite{hylton_reduced_1984, MOHR1998227} for further details. In order to find the explicit form of these functions, we have to specify the interaction potential between the leptons and the nucleus. By assuming Delbrück scattering on a bare point-like nucleus, the functions $F^{1,2}_{\kappa,0}(r)$ and $F^{1,2}_{\kappa,\infty}(r)$ are given by

\begin{widetext}
\begin{equation} \label{GreenCoulomb}
\begin{aligned}
\left[\begin{array}{c}
F^1_{\kappa,0}(x,z)\\
F^2_{\kappa,0}(x,z)
\end{array}\right] &= \left[\begin{array}{c}
\frac{\sqrt{1+z}}{2cx^{3/2}} \left((\lambda-\nu)M_{\nu-1/2,\lambda}(2cx) - \left(\kappa-\frac{\gamma}{c} \right)M_{\nu+1/2,\lambda}(2cx)\right)\\
\frac{\sqrt{1-z}}{2cx^{3/2}} \left((\lambda-\nu)M_{\nu-1/2,\lambda}(2cx) + \left(\kappa-\frac{\gamma}{c} \right)M_{\nu+1/2,\lambda}(2cx)\right)\\
\end{array}\right]~,\\
\left[\begin{array}{c}
F^1_{\kappa,\infty}(x,z)\\
F^2_{\kappa,\infty}(x,z)
\end{array}\right] &= \frac{\Gamma(\lambda-\nu)}{\Gamma(1+2\lambda)}\left[\begin{array}{c}
\frac{\sqrt{1+z}}{2cx^{3/2}} \left(\left(\kappa+\frac{\gamma}{c} \right)W_{\nu-1/2,\lambda}(2cx) + W_{\nu+1/2,\lambda}(2cx)\right)\\
\frac{\sqrt{1-z}}{2cx^{3/2}} \left(\left(\kappa+\frac{\gamma}{c} \right)W_{\nu-1/2,\lambda}(2cx) - W_{\nu+1/2,\lambda}(2cx)\right)\\
\end{array}\right]~,
\end{aligned}
\end{equation}
\end{widetext}

\noindent where $M_{\kappa,\mu}$ and $W_{\kappa,\mu}$ are the Whittaker functions and $c = \sqrt{1-z^2}$, $\gamma = \alpha Z$, $\nu = \gamma z/c$ and $\lambda = \sqrt{\kappa^2-\gamma^2}$~\cite{MOHR1998227}. Here, the branch of the square root is taken so that Re$(c) \geq 0$ and, moreover, the Wronskian~\eqref{Wronskian} is just unity, $w_\kappa(z) = 1$.

Having derived the components of the radial Green's function, we can obtain the full function~\eqref{GreenSpec}, that enters the Delbrück scattering matrix element, as

\begin{widetext}
\begin{equation} \label{AnalyticalGreen}
\begin{aligned}
G(&\boldsymbol{r}_2,\boldsymbol{r}_1,z) = \sum_{\kappa\mu} \frac{1}{w_\kappa(z)} \Bigg[\Theta(r_2-r_1) \left(\begin{array}{c}
F_{\kappa,\infty}^1 (r_2,z) \chi_\kappa^\mu(\hat{\boldsymbol{r}}_2)\\
iF_{\kappa,\infty}^2 (r_2,z) \chi_{-\kappa}^\mu(\hat{\boldsymbol{r}}_2)
\end{array}\right)
\left(\begin{array}{cc}
F_{\kappa,0}^1 (r_1,z) \chi_\kappa^{\mu\dagger}(\hat{\boldsymbol{r}}_1) & -iF_{\kappa,0}^2 (r_1,z) \chi_{-\kappa}^{\mu\dagger}(\hat{\boldsymbol{r}}_1)
\end{array}\right)\\
&+ \Theta(r_1-r_2) \left(\begin{array}{c}
F_{\kappa,0}^1 (r_2,z) \chi_\kappa^\mu(\hat{\boldsymbol{r}}_2)\\
iF_{\kappa,0}^2 (r_2,z) \chi_{-\kappa}^\mu(\hat{\boldsymbol{r}}_2)
\end{array}\right)
\left(\begin{array}{cc}
F_{\kappa,\infty}^1 (r_1,z) \chi_\kappa^{\mu\dagger}(\hat{\boldsymbol{r}}_1) & -iF_{\kappa,\infty}^2 (r_1,z) \chi_{-\kappa}^{\mu\dagger}(\hat{\boldsymbol{r}}_1)
\end{array}\right)\Bigg]\\
&\equiv \sum_{\kappa\mu} \frac{1}{w_\kappa(z)} \left[\Theta(r_2-r_1) \mathcal{F}_{\kappa,\infty}^\mu(\boldsymbol{r}_2,z) \tilde{\mathcal{F}}_{\kappa,0}^\mu(\boldsymbol{r}_1,z) + \Theta(r_1-r_2) \mathcal{F}_{\kappa,0}^\mu(\boldsymbol{r}_2,z) \tilde{\mathcal{F}}_{\kappa,\infty}^\mu(\boldsymbol{r}_1,z) \right]~.
\end{aligned}
\end{equation}
\end{widetext}

\noindent In the last line of this expression, we introduced the short-hand notation $\mathcal{F}_{\kappa,\infty}^\mu$, $\tilde{\mathcal{F}}_{\kappa,0}^\mu$, $\mathcal{F}_{\kappa,0}^\mu$ and $\tilde{\mathcal{F}}_{\kappa,\infty}^\mu$ to represent the row and column spinors from which the Green's function is constructed.

\subsection{Evaluation of the matrix element} \label{EvMatrix}
In the previous two sections, we obtained the radial-angular decomposition of the photon-lepton interaction operator (\ref{ElPhOp}) and the Green's function (\ref{AnalyticalGreen}). By inserting these expansions into Eq. \eqref{MatrixElement}, we can write the Delbrück scattering amplitude as

\begin{equation} \label{AmplMult}
\begin{aligned}
&M_{\lambda_1,\lambda_2} =  i\alpha \sum_{\kappa\mu}\sum_{\kappa'\mu'}\sum_{P_1L_1M_1}\sum_{P_2L_2M_2}i^{L_1+P_1-L_2-P_2}\\
&\hspace{0.5cm}\times\lambda_1^{P_1}\lambda_2^{P_2}\sqrt{(2L_1+1)(2L_2+1)}D^{L_1}_{M_1\lambda_1}(\phi_1,\theta_1,0)\\
&\hspace{0.5cm}\times D^{L_2*}_{M_2\lambda_2}(\phi_2,\theta_2,0)\Big[I_1(\kappa'\mu',\omega,P_1L_1M_1,P_2L_2M_2,\kappa\mu)\\
&\hspace{0.5cm}+I_2(\kappa'\mu',\omega,P_1L_1M_1,P_2L_2M_2,\kappa\mu)\Big]~,\\
\end{aligned}
\end{equation}

\noindent where we now explicitly show the dependence on the helicity of the incoming and outgoing photon $\lambda_1$ and $\lambda_2$. The non-trivial part of the calculations is contained within the multi-dimensional integrals

\begin{equation} \label{I1}
\begin{aligned}
&I_1(\kappa'\mu',\omega,P_1L_1M_1,P_2L_2M_2,\kappa\mu) \\
&=\int_{-\infty}^\infty\text{d}z'~\int_{-\infty}^\infty\text{d}z~\frac{\delta (\omega+z-z')}{w_\kappa(z)w_{\kappa'}(z')}\\
&\times\int \text{d}^3\boldsymbol{r}_1~\tilde{\mathcal{F}}_{\kappa',\infty}^{\mu'}(\boldsymbol{r}_1,z')\boldsymbol{\alpha}\cdot\boldsymbol{a}_{L_1M_1}^{(P_1)}(\boldsymbol{r}_1) \mathcal{F}_{\kappa,\infty}^\mu(\boldsymbol{r}_1,z)\\
&\times \left[\int_{r_2 \leq r_1} \text{d}^3\boldsymbol{r}_2~ \mathcal{F}_{\kappa',0}^{\mu'}(\boldsymbol{r}_2,z')^\dagger \boldsymbol{\alpha}\cdot\boldsymbol{a}_{L_2M_2}^{(P_2)}(\boldsymbol{r}_2)\tilde{\mathcal{F}}_{\kappa,0}^\mu(\boldsymbol{r}_2,z)^\dagger\right]^*\\
\end{aligned}
\end{equation}

\noindent and

\begin{equation} \label{I2}
\begin{aligned}
&I_2(\kappa'\mu',\omega,P_1L_1M_1,P_2L_2M_2,\kappa\mu)\\
&= \int_{-\infty}^\infty\text{d}z'~\int_{-\infty}^\infty\text{d}z~\frac{\delta (\omega+z-z')}{w_\kappa(z)w_{\kappa'}(z')}\\
&\times \left[\int \text{d}^3\boldsymbol{r}_2~\mathcal{F}_{\kappa',\infty}^{\mu'}(\boldsymbol{r}_2,z')^\dagger\boldsymbol{\alpha}\cdot\boldsymbol{a}_{L_2M_2}^{(P_2)}(\boldsymbol{r}_2)\tilde{\mathcal{F}}_{\kappa,\infty}^\mu(\boldsymbol{r}_2,z)^\dagger\right]^*\\
&\times \int_{r_1 \leq r_2} \text{d}^3\boldsymbol{r}_1~\tilde{\mathcal{F}}_{\kappa',0}^{\mu'}(\boldsymbol{r}_1,z')\boldsymbol{\alpha}\cdot\boldsymbol{a}_{L_1M_1}^{(P_1)}(\boldsymbol{r}_1) \mathcal{F}_{\kappa,0}^\mu(\boldsymbol{r}_1,z)~.\\
\end{aligned}
\end{equation}

\noindent These functions can be represented as the product of radial and angular integrals, see Appendix A. While the angular integrals can be easily calculated using Racah algebra \cite{10.5555/1215645, Grant_1974}, the radial counterparts require some more attention.

\section{Evaluation of the Delbrück amplitude} \label{EvInt}
\subsection{Divergence} \label{renorm}
In the previous section we obtained expressions for the scattering amplitude represented by the Feynman diagram in Fig.~1. These expressions cannot be directly used for a numerical evaluation since the integration over the intermediate-state energies is divergent in the ultraviolet region. In order to eliminate this divergency, it is convenient to consider the expansion of the electron propagators in the vacuum-polarization loop in terms of interactions with the binding nuclear field, i.e., in powers of the coupling constant
$(\alpha Z)$. The first term of the expansion contains only the free electron propagators and does not contribute to the Delbr\"uck scattering since it does not involve the nucleus. This $Z$-independent term is conveniently eliminated by subtracting the $Z = 0$ contribution before
integrations, see Fig.~2.  Similar methods have been used many times in the past for Delbr\"uck scattering~\cite{PhysRevD.45.2982,scherdin_coulomb_1995} and for the nonperturbative -- to all orders in $\alpha Z$ -- calculation of the vacuum polarization diagram, e.g. Refs.~\cite{GYULASSY1975497, PhysRevA.38.5066, PhysRevA.48.2772, PhysRevA.56.3529, PhysRevA.89.042121}.

The second term of the potential expansion of the vacuum-polarization loop is linear in $Z$. This term (as well as all other odd-$Z$ contributions) vanish due to Furry's theorem. However, such terms are present in the integrand and may lead to spurious contributions. We eliminate such terms by expressing the integrand to be symmetrical with respect to $Z\to -Z$.

The higher-order terms of the potential expansion contain two or more interactions with the nuclear binding field; they are ultraviolet finite. However, it is known~\cite{GYULASSY1975497, PhysRevA.12.748} that the vacuum-polarization loop with four vertices might induce spurious gauge-noninvariant contributions. In the present work we use the demonstration of Refs.~\cite{PhysRevA.56.3529, PhysRevA.60.45} that for a finite energy transferred through the loop, the spurious terms vanish when the partial-wave summation is performed after all integrations. This prescription corresponds to our scheme of calculations, so we conclude that no spurious terms arise in our approach.

\begin{figure}
\begin{center}
\includegraphics[width=0.85\linewidth]{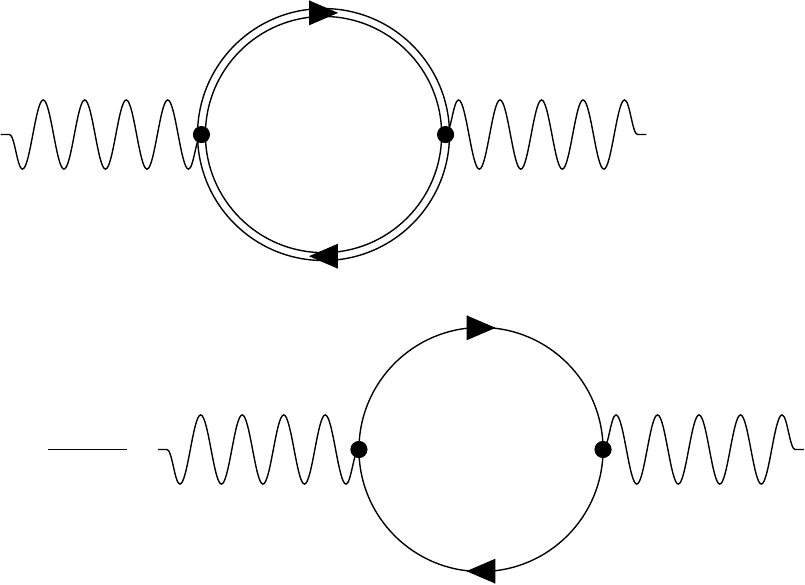}\caption{Subtraction of the divergent free-loop contribution from the full Delbrück scattering Feynman diagram.} \label{FeynmanSub}
\end{center}
\end{figure}
\subsection{Energy integration} \label{EInt}
\begin{figure}
\begin{center}
\includegraphics[width=0.85\linewidth]{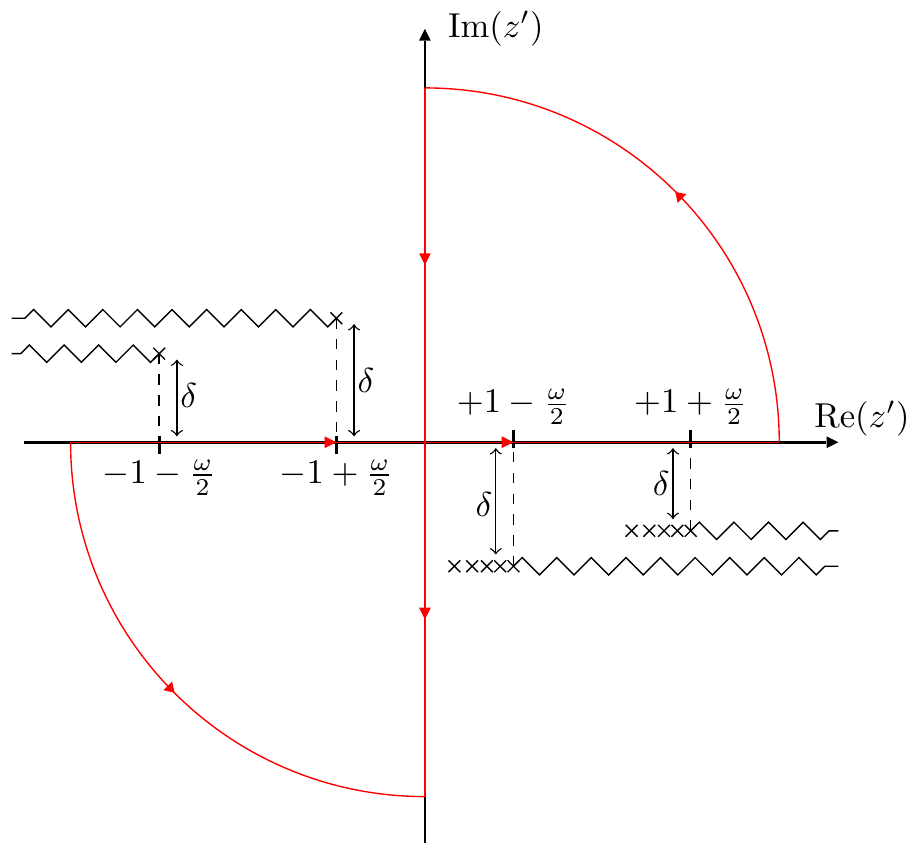}
\caption{Schematic depiction of the singularities (black crosses), branch cuts (black zig zag lines) and infinitesimal shift of the poles from the real axis $\delta$. The red line shows the closed integral over $z'$ in $K_{ij}$ which vanishes due to Cauchy's integral formula.} \label{IntPath}
\end{center}
\end{figure}

After extracting the free-loop contribution from the Delbrück scattering diagram, we have to numerically calculate the remaining part of the amplitude. As seen from Eq.~\eqref{AmplMult}, this requires the evaluation of the functions $I_1$ and $I_2$ which contain the integration over the energies $z$ and $z'$. The integration over $z$ can be performed trivially owing to the Dirac delta function $\delta (\omega + z - z')$ that gives $z = z'-\omega$. The remaining integration over $z'$ has to be performed numerically. The stability of this numerical integration can be improved by substituting $z' \to z' + \tfrac{\omega}{2}$ thus making the integrand symmetric with respect to the origin. This symmetry results in the integrand being invariant to charge inversion $Z \to -Z$ meaning that all contributions proportional to odd powers of $Z$ vanish before the integration over $z'$. The odd-$Z$ contributions being cancelled out makes the calculation more stable. The resulting integrand is analytical in the entire complex plane except for the branch cuts starting at $z' = \pm 1 + \frac{\omega}{2}$ and $z' = \pm 1 - \frac{\omega}{2}$ as well as the poles at $z' = (\lambda' + m)/\sqrt{\gamma^2 + (\lambda'+m)^2} + \frac{\omega}{2}$ and $z' = (\lambda + m)/\sqrt{\gamma^2 + (\lambda+m)^2} - \frac{\omega}{2}$, $m = 0,1,2,...$, see Fig. \ref{IntPath}. Since these poles are located infinitely close to the real axis, $\delta \to 0$, the energy integration in the interval $(-\infty, \infty)$ might be troublesome. To avoid this, let us first assume that the photon energy is under the pair creation threshold, $\omega/2 < 1- e_0$, where $e_0$ is the ionization energy of the $1s$ Dirac state. In this case, the analytical structure of the integrand is shown in Fig.~\ref{IntPath} and we can perform the well-known Wick rotation of the integration contour

\begin{equation}
\int_{-\infty}^\infty \text{d}z' \to \int_{-i\infty}^{i\infty} \text{d}z'~,
\end{equation}

\noindent in which we turn the integration path to the imaginary axis. By using Cauchy's integral formula, it can be shown that the integral along the rotated contour is equivalent to the original integral along the real axis. The integration contour as displayed in Fig.~\ref{IntPath} allows us to perform the integral over $z'$ in a stable and controlled way. 

\subsection{Radial integration} \label{radInt}
Having discussed the energy integration of the functions \eqref{I1} and \eqref{I2}, we are ready now to address the remaining angular and radial integrals. As mentioned in section \ref{EvMatrix}, the angular integrals can be easily calculated analytically using Racah algebra while their radial counterparts are more complicated. In appendix~A, we show that the main building blocks for the radial part are the integrals

\begin{equation} \label{IntI}
\begin{aligned}
\mathcal{J}&(z',\kappa',\omega,L_1,L_2,\kappa,p_1,p_2,p_3,p_4) \\
&= \frac{\Gamma(\lambda-\nu)\Gamma(\lambda'-\nu')}{\Gamma(1+2\lambda)\Gamma(1+2\lambda')}\\
&\times\int_0^\infty \frac{\text{d}r_1}{r_1} W_{\nu'+\frac{p_1}{2},\lambda'}(2c'r_1) j_{L_1}(\omega r_1) W_{\nu+\frac{p_2}{2},\lambda}(2cr_1)\\
&\times\int_0^{r_1} \frac{\text{d}r_2}{r_2}M_{\nu'+\frac{p_3}{2},\lambda'}(2c'r_2) j_{L_2}(\omega r_2) M_{\nu+\frac{p_4}{2},\lambda}(2cr_2)~,\\
\end{aligned}
\end{equation}

\noindent where $L_1$ and $L_2$ are the multiplicities of the incoming and outgoing photon, $\kappa'$ and $\kappa$ are the Dirac quantum numbers of the electron and positron propagators and $p_1$,...,$p_4$ are parameters that can take the values $+1$ or $-1$. The calculation of these integrals is a computationally demanding task. One of the main reasons for this is the fact that the integrand is a strongly oscillating function. Moreover, owing to the properties of the Whittaker functions, the amplitude of this oscillation decreases polynomially and not exponentially with the radial coordinate $r_1$. Hence, to achieve a good convergence of the results, the numerical calculation has to be performed with many integration points and within a large interval, leading to long computation times. In order to overcome this difficulty, we apply an approach in which the \textit{numerical} integration runs from $r_1 = 0$ to some arbitrary parameter $a$ while the remaining integration from $r_1 = a$ to infinity is done \textit{analytically}

\begin{widetext}
\begin{equation} \label{IntSplit}
\begin{aligned}
\mathcal{J}&(z',\kappa',\omega,L_1,L_2,\kappa,p_1,p_2,p_3,p_4) = \frac{\Gamma(\lambda-\nu)\Gamma(\lambda'-\nu')}{\Gamma(1+2\lambda)\Gamma(1+2\lambda')}\\
&\times\Bigg[ \underbrace{\int_0^a \frac{\text{d}r_1}{r_1} W_{\nu'+\frac{p_1}{2},\lambda'}(2c'r_1) j_{L_1}(\omega r_1) W_{\nu+\frac{p_2}{2},\lambda}(2cr_1)\int_0^{r_1} \frac{\text{d}r_2}{r_2} M_{\nu'+\frac{p_3}{2},\lambda'}(2c'r_2) j_{L_2}(\omega r_2) M_{\nu+\frac{p_4}{2},\lambda}(2cr_2)}_\text{Numerical integration}\\
&+\underbrace{\int_a^\infty \frac{\text{d}r_1}{r_1} W_{\nu'+\frac{p_1}{2},\lambda'}(2c'r_1) j_{L_1}(\omega r_1) W_{\nu+\frac{p_2}{2},\lambda}(2cr_1)}_\text{Analytical integration}\underbrace{\int_0^{a'} \frac{\text{d}r_2}{r_2} M_{\nu'+\frac{p_3}{2},\lambda'}(2c'r_2) j_{L_2}(\omega r_2) M_{\nu+\frac{p_4}{2},\lambda}(2cr_2)}_\text{Numerical integration}\\
&+\underbrace{\int_a^\infty \frac{\text{d}r_1}{r_1} W_{\nu'+\frac{p_1}{2},\lambda'}(2c'r_1) j_{L_1}(\omega r_1) W_{\nu+\frac{p_2}{2},\lambda}(2cr_1)\int_{a'}^{r_1} \frac{\text{d}r_2}{r_2} M_{\nu'+\frac{p_3}{2},\lambda'}(2c'r_2) j_{L_2}(\omega r_2) M_{\nu+\frac{p_4}{2},\lambda}(2cr_2)}_\text{Analytical integration}\Bigg]~.
\end{aligned}
\end{equation}
\end{widetext}

Such an approach is possible due to the well-known  asymptotic expansions of the Whittaker functions for large arguments

\begin{subequations} \label{asympWhit}
\begin{align}
\begin{split}
M_{\alpha,\beta}(z) &\sim \frac{\Gamma(1+2\beta)}{\Gamma(\frac{1}{2}+\beta-\alpha)} e^{\frac{1}{2}z}z^{-\alpha}\sum_{s=0}^\infty u_M(s,\alpha,\beta)z^{-s}\\
&+ \frac{\Gamma(1+2\beta)}{\Gamma(\frac{1}{2}+\beta+\alpha)} e^{-\frac{1}{2}z\pm(\frac{1}{2}+\beta-\alpha)\pi i}z^{\alpha}\\
&\times\sum_{s=0}^\infty \widetilde{u}_M(s,\alpha,\beta)(-z)^{-s}~,\\
\end{split}\\
&W_{\alpha,\beta}(z) \sim e^{-\frac{1}{2}z}z^{\alpha}\sum_{s=0}^\infty u_W(s,\alpha,\beta)(-z)^{-s}~,
\end{align}
\end{subequations}

\noindent where

\begin{subequations}
\begin{align}
u_M(s,\alpha,\beta) &= \frac{(\frac{1}{2}-\beta+\alpha)_s(\frac{1}{2}+\beta+\alpha)_s}{s!}~,\\
\widetilde{u}_M(s,\alpha,\beta) &= \frac{(\frac{1}{2}+\beta-\alpha)_s(\frac{1}{2}-\beta-\alpha)_s}{s!}~,\\
u_W(s,\alpha,\beta) &= \frac{(\frac{1}{2}+\beta-\alpha)_s(\frac{1}{2}-\beta-\alpha)_s}{s!}~,
\end{align}
\end{subequations}

\noindent and $(x)_n$ is the Pochhammer symbol. Moreover, we use the exact expansion of the spherical Bessel function which reads

\begin{equation} \label{asympBes}
\begin{aligned}
j_L(x) &= \sum_{m=0}^L \frac{(L+m)!}{m!(L-m)!}i^{L+1-m}(2x)^{-m-1}\\
&\times[(-1)^{L+1-m}e^{ix}+e^{-ix}]~,
\end{aligned}
\end{equation}

\noindent see Ref.~\cite{Abramowitz_1974}. We choose $a$ and $a'$ in Eq.~\eqref{IntSplit} large enough so that the asymptotic expansion of the Whittaker functions above converges fast. By inserting the expressions \eqref{asympWhit} - \eqref{asympBes} into the third and fourth line of Eq. \eqref{IntSplit}, we can perform the analytical integration up to all orders in the asymptotic expansions, see Appendix B.

\subsection{Computational details} \label{compDet}
We have discussed above the method for the calculation of Delbrück scattering amplitudes. However, the practical implementation of this method still comes with some difficulties that need to be addressed in order to keep the numerical uncertainties under control. The main source of numerical error comes from the summation over the multipole components $\kappa'$, $\kappa$, $L_1$ and $L_2$. It can be easily seen that for fixed $\kappa'$ and $\kappa$, the sum over the photon multipole components $L_1$ and $L_2$ is bound by the properties of the angular integrals. In practice, it is usually not necessary to carry out the summation over all non-vanishing multipoles because for most scattering angles, a satisfactory convergence is reached earlier. In contrast to the summation over $L_1$ and $L_2$, the sum over $\kappa'$ and $\kappa$ is infinite and converges very slowly. In order to accelerate this convergence, we utilize a trick that is based on the fact that the full Delbrück amplitude vanishes for zero photon energy when all integrations and summations have been carried out. Therefore, subtracting the amplitude for $\omega = 0$ does not change the final result. However, we found that subtracting this amplitude before carrying out the summation over $\kappa'$ and $\kappa$ vastly increases the convergence and we can reduce the number of needed multipoles significantly by using this trick. The number of combinations of $\kappa'$ and $\kappa$ that need to be calculated can be further minimized by using an extrapolation technique. The application of this technique requires us to first write the sum as

\begin{equation} \label{sumKappa}
\sum_{\kappa,\kappa'} X_{\kappa,\kappa'} = \sum_{|\kappa|,|\kappa'|} X_{|\kappa|,|\kappa'|} = \sum_{|\kappa|} Y_{|\kappa|}~,
\end{equation}

\noindent where we have introduced

\begin{equation}
X_{|\kappa|,|\kappa'|} = \sum_{\text{sign}(\kappa)} \sum_{\text{sign}(\kappa')}X_{\kappa,\kappa'}~,
\end{equation}

\noindent and

\begin{equation}
Y_{|\kappa|} = \sum_{|\kappa'|=1}^{|\kappa|}(X_{|\kappa|,|\kappa'|} + X_{|\kappa'|,|\kappa|}) - X_{|\kappa|,|\kappa|}~.
\end{equation}

\noindent We see from Eq.~\eqref{sumKappa} that the summation over $\kappa'$ and $\kappa$ has been written as a sum over a single parameter $|\kappa|$. We now evaluate the sum over $Y_{|\kappa|}$ up to some maximum value $|\kappa|_\text{max}$ and estimate the tail of the expansion using least-squares inverse polynomial fitting. Utilizing this method, we achieve a relative accuracy of less than 1\% for scattering angles $\theta > 30^\circ$ and less than 3\% for $\theta \leq 30^\circ$ by summing up to $|\kappa|_\text{max} = 40$ for all calculations shown in the next section.

In contrast to the multipole expansions discussed above, the other parts of the calculations don't introduce such large uncertainties into the results. However, to keep the respective errors small, some further numerical tricks are required. For example, both, the evaluation of the complex Whittaker functions in Eq.~\eqref{IntI} and the analytical solution of the asymptotic integrals in Eq.~\eqref{IntSplit}, suffer from severe numerical cancellations as well as over- and underflow problems. Furthermore, the subtraction of the free loop contribution in Fig.~\ref{FeynmanSub} can also cancel up to nine digits of accuracy. Hence, performing calculations with double precision arithmetics could lead to the loss of all significant digits. To solve the numerical issues described above, we use arbitrary precision ball point arithmetics, as implemented by Johansson~\cite{johansson_arb:_2017}, in our entire integration routine to get a high precision and rigorous error bounds for the integrals. The working precision is chosen large enough so that rounding errors are negligible for our calculations.

In addition to the cancellation problems, another source of numerical uncertainty originates from the integration over $r_1$, $r_2$ and $z'$ in Eqs.~\eqref{I1} and \eqref{I2}. This uncertainty is mainly caused by the numerical integration itself, which we perform using Gauß-Legendre quadrature~\cite{Press_2007}, and the summation over the asymptotic expansion of the Whittaker functions~\eqref{asympWhit}, which has to be terminated at some summation index~$s$. In all calculations shown in this work, we choose the number of integration points and the number of terms in the summation over $s$ large enough, so that the relative numerical error in the final amplitude is smaller than $10^{-6}$ and, therefore, also negligible. In order to avoid the introduction of any additional error by truncating the integral over $z'$, which goes up to infinity, see Sec.~\ref{EInt}, we split the integral in each half of the complex plane into two parts. First we integrate from zero to some arbitrary parameter $A$ and then we integrate the remaining part up to infinity by mapping it to the interval $[0,1]$ through the change of variable $z' = A/t$.

To be able to run our calculations in a reasonable time frame, the evaluation of the radial integrals for different $z'$ is done in parallel on the PTB high performance cluster. A typical calculation runs on 72 threads simultaneously, which vastly increases the performance and reduces the total computing time to a approximately two days for a full set of amplitudes, including all scattering angles and photon polarizations.

\section{Numerical Results} \label{numResults}

\begin{figure*}
    \centering
      \includegraphics[width=0.85\textwidth]{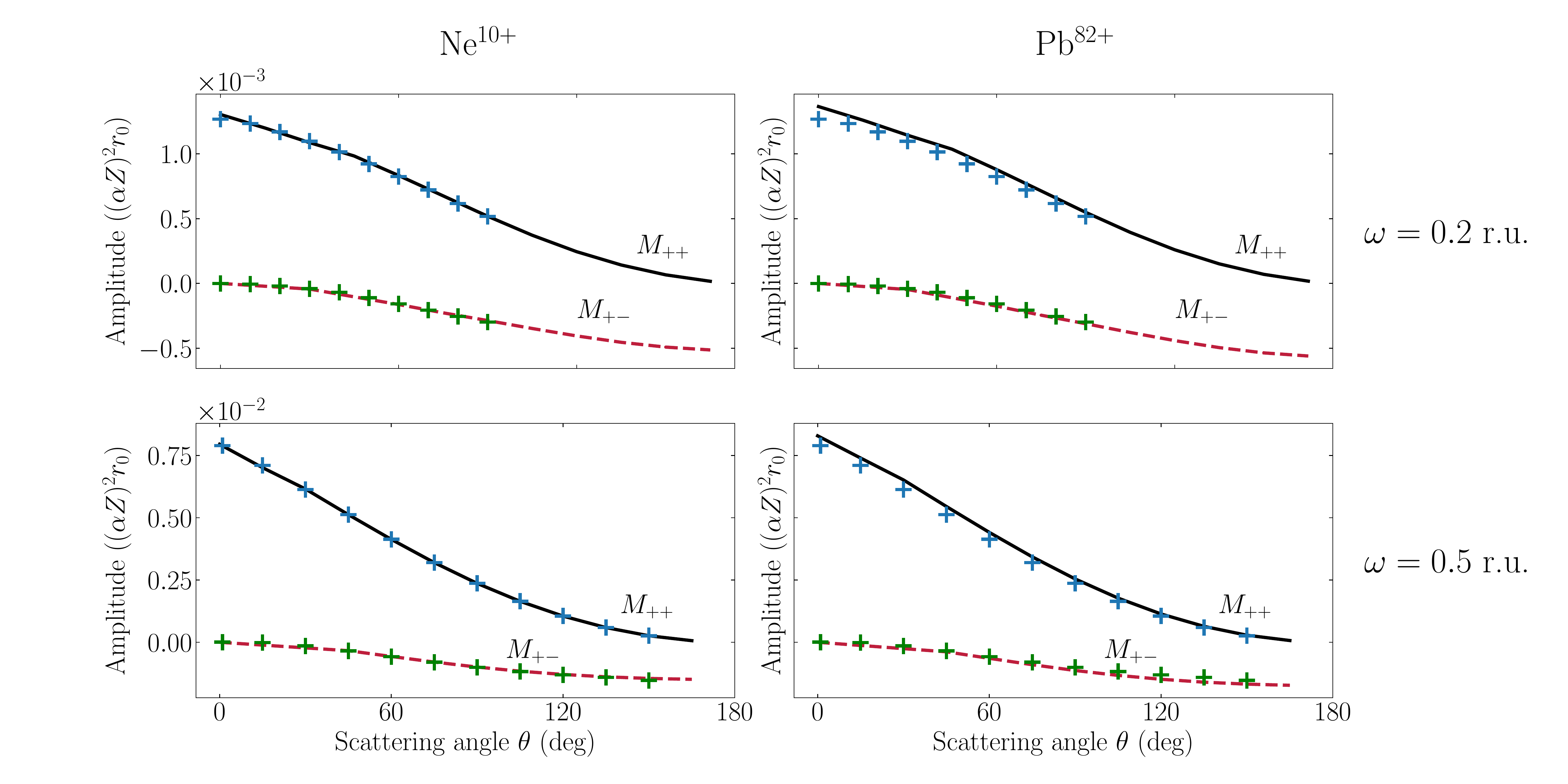}
       \caption{Amplitudes for Delbrück scattering~\eqref{AmplMult} in collisions of photons with energies $\omega = 0.2~\text{r.u.}~(102.2~\text{keV})$ (upper panels) and $\omega = 0.5~\text{r.u.}~(255.5~\text{keV})$ (lower panels) with bare neon (left panels) and lead (right panels) nuclei in units $(\alpha Z)^2r_0$, where $r_0 = 2.818~\text{fm}$ is the classical electron radius. Calculations have been performed for the non-helicity flip amplitude $M_{++} = M_{--}$ (black solid line) as well as the helicity flip amplitude $M_{+-} = M_{-+}$ (red dashed line). Moreover, we show the lowest-order Born approximation (crosses) as given by Ref. \cite{PhysRevD.12.206} (upper panel) and \cite{FALKENBERG19921} (lower panel). \label{Amplitdues}}
\end{figure*}

In the previous sections, we have established the theoretical framework to calculate amplitudes for Delbrück scattering. Moreover, we have discussed an efficient method to evaluate the multidimensional integrals that are involved in this calculation. In what follows, we show the viability of our method by presenting numerical results. In Fig. \ref{Amplitdues}, we display the Delbrück scattering amplitude~\eqref{AmplMult} for the collision of photons with energies $\omega = 0.2~\text{r.u.}~(102.2~\text{keV})$ and $\omega = 0.5~\text{r.u.}~(255.5~\text{keV})$ with bare neon and lead nuclei. For each scenario, we investigate the scattering in which the photon helicity is either flipped, $\lambda_1 \neq \lambda_2$, or unchanged, $\lambda_1 = \lambda_2$. The corresponding amplitudes are conveniently referred to as helicity flip and non-helicity flip amplitudes and are denoted as $M_{++} = M_{--}$ and $M_{+-} = M_{-+}$. The fact that out of four amplitudes, only two are independent is well-known and can be explained by symmetry reasons~\cite{MILSTEIN1994183}. We compare our numerical results with the lowest-order Born approximation as given by Ref.~\cite{FALKENBERG19921, PhysRevD.12.206}. Following the convention of Falkenberg and co-workers~\cite{FALKENBERG19921}, we present our results in units $(\alpha Z)^2r_0$, where $r_0 = 2.818~\text{fm}$ is the classical electron radius.

In the left column of the figure, we compare our results with the predictions of the lowest-order Born approximation for a neon target. The Born approximation can be obtained from an $\alpha Z$ expansion of the Feynman diagram in Fig. \ref{FeynmanDel} and neglecting all terms of higher order than $(\alpha Z)^2$. As seen from the left column of the figure, the Born and our rigorous results agree very well, as can be expected for the low-$Z$ regime. Moreover, as seen from the right column, the normalized amplitudes are almost unchanged for the high-$Z$ regime and the higher order corrections mainly lead to a slight enhancement of the absolute value of the scattering amplitude in the order of a few percent. By comparing the upper and lower panels of the figure, we can observe, moreover, the well-known $\omega^2$ low-energy scaling of the Delbrück amplitude~\cite{PhysRevD.12.206}.

\section{Summary and Outlook} \label{sumAOut}
In conclusion, we presented a theoretical method to calculate amplitudes for Delbrück scattering, which accounts for the interaction of the virtual electron-positron pairs with the nucleus up to all orders. A special emphasis was put on the practical evaluation of the multidimensional integrals that are involved in this calculation. In particular, we found an analytical solution of the radial integrals up to all orders in the asymptotic expansions of the involved special functions. By using a combination of numerical and analytical integration methods, we were able to improve the numerical stability of our calculations significantly and compute Delbrück amplitudes in a well-controlled way. In order to illustrate the use of our proposed method, we have performed calculations of relatively low energy photons colliding with bare neon and lead ions. As expected, the results of these calculations have been found in very good agreement with the lowest-order Born approximation. Our results suggest that the Coulomb corrections lead to an enhancement of the absolute value of the scattering amplitude of about a few percent for all scattering angles and relatively low energies.

While higher-order corrections are relatively small for the parameter regime studied in this work, it is well-known, both, from theoretical analysis~\cite{PhysRev.108.169} and experiments~\cite{rullhusen_coulomb_1979}, that these effects are much larger above the pair production threshold. The method developed in this work can be applied to these energies with a few modifications to the integration path and some more optimizations to decrease the computation time. A study of Delbrück scattering above the pair production threshold is currently under development and will be published in a future work.

\begin{acknowledgments}
This work has been supported by the GSI Helmholtz Centre for Heavy Ion Research under the project BSSURZ1922. A.V.V. acknowledges financial support by the Government of the Russian Federation through the ITMO Fellowship and Professorship Program. We would also like to thank our colleagues from PTB's high performance computing division and especially Gert Lindner for providing access to their computation cluster and for their excellent technical support.
\end{acknowledgments}

\appendix
\begin{widetext}
\section{Evaluation of the functions $I_1$ and $I_2$}
In this section, we want to express the integrals in Eqs.~\eqref{I1} and \eqref{I2} in terms of their radial and angular parts. We start by rewriting the functions $I_1$ and $I_2$ as

\begin{equation} \label{I1App}
\begin{aligned}
&I_1(\kappa'\mu',\omega,P_1L_1M_1,P_2L_2M_2,\kappa\mu) =\int_{-\infty}^\infty\text{d}z'~\int_{-\infty}^\infty\text{d}z~\frac{\delta (\omega+z-z')}{w_\kappa(z)w_{\kappa'}(z')}\\
&\times\int \text{d}^3\boldsymbol{r}_1~\tilde{F}_{\kappa',\infty}^{\mu'}(\boldsymbol{r}_1,z')\boldsymbol{\alpha}\cdot\boldsymbol{a}_{L_1M_1}^{(P_1)}(\boldsymbol{r}_1) F_{\kappa,\infty}^\mu(\boldsymbol{r}_1,z) \left[\int_{r_2 \leq r_1} \text{d}^3\boldsymbol{r}_2~ F_{\kappa',0}^{\mu'}(\boldsymbol{r}_2,z')^\dagger \boldsymbol{\alpha}\cdot\boldsymbol{a}_{L_2M_2}^{(P_2)}(\boldsymbol{r}_2)\tilde{F}_{\kappa,0}^\mu(\boldsymbol{r}_2,z)^\dagger\right]^*\\
&\equiv \int_{-\infty}^\infty\text{d}z'~\int_{-\infty}^\infty\text{d}z~\frac{\delta (\omega+z-z')}{w_\kappa(z)w_{\kappa'}(z')}\int \text{d}^3\boldsymbol{r}_1~\tilde{F}_{\kappa',\infty}^{\mu'}(\boldsymbol{r}_1,z')\boldsymbol{\alpha}\cdot\boldsymbol{a}_{L_1M_1}^{(P_1)}(\boldsymbol{r}_1) F_{\kappa,\infty}^\mu(\boldsymbol{r}_1,z) J_1^*(\omega,z'\kappa'\mu',P_2L_2M_2,z\kappa\mu, r_1)~,\\
\end{aligned}
\end{equation}

\noindent and

\begin{equation}
\begin{aligned}
&I_2(\kappa'\mu',\omega,P_1L_1M_1,P_2L_2M_2,\kappa\mu)= \int_{-\infty}^\infty\text{d}z'~\int_{-\infty}^\infty\text{d}z~\frac{\delta (\omega+z-z')}{w_\kappa(z)w_{\kappa'}(z')}\\
&\times \left[\int \text{d}^3\boldsymbol{r}_2~F_{\kappa',\infty}^{\mu'}(\boldsymbol{r}_2,z')^\dagger\boldsymbol{\alpha}\cdot\boldsymbol{a}_{L_2M_2}^{(P_2)}(\boldsymbol{r}_2)\tilde{F}_{\kappa,\infty}^\mu(\boldsymbol{r}_2,z)^\dagger\right]^* \int_{r_1 \leq r_2} \text{d}^3\boldsymbol{r}_1~\tilde{F}_{\kappa',0}^{\mu'}(\boldsymbol{r}_1,z')\boldsymbol{\alpha}\cdot\boldsymbol{a}_{L_1M_1}^{(P_1)}(\boldsymbol{r}_1) F_{\kappa,0}^\mu(\boldsymbol{r}_1,z)\\
&\equiv \int_{-\infty}^\infty\text{d}z'~\int_{-\infty}^\infty\text{d}z~\frac{\delta (\omega+z-z')}{w_\kappa(z)w_{\kappa'}(z')} \left[\int \text{d}^3\boldsymbol{r}_2~F_{\kappa',\infty}^{\mu'}(\boldsymbol{r}_2,z')^\dagger\boldsymbol{\alpha}\cdot\boldsymbol{a}_{L_2M_2}^{(P_2)}(\boldsymbol{r}_2)\tilde{F}_{\kappa,\infty}^\mu(\boldsymbol{r}_2,z)^\dagger\right]^* J_2(\omega,z'\kappa'\mu',P_1L_1M_1,z\kappa\mu, r_2)~.
\end{aligned}
\end{equation}

\noindent To compute the involved integrals efficiently, we have to evaluate these expressions further. Let us start by writing $J_1$ as

\begin{equation}
\begin{aligned}
J_1(z'\kappa'\mu',\omega,P_2L_2M_2,z\kappa\mu, r_1) = i\int_{r_2 \leq r_1} \text{d}^3\boldsymbol{r}_2~ [&F_{\kappa',0}^{1}(r_2,z')^*(\chi_{\kappa'}^{\mu'})^\dagger(\hat{\boldsymbol{r}}_2)\boldsymbol{\sigma}\cdot \boldsymbol{a}_{L_2M_2}^{(P_2)}(\boldsymbol{r}_2)F_{\kappa,0}^{2}(r_2,z)^*\chi_{-\kappa}^{\mu}(\hat{\boldsymbol{r}}_2)\\
&-F_{\kappa',0}^{2}(r_2,z')^*(\chi_{-\kappa'}^{\mu'})^\dagger(\hat{\boldsymbol{r}}_2)\boldsymbol{\sigma}\cdot \boldsymbol{a}_{L_2M_2}^{(P_2)}(\boldsymbol{r}_2)F_{\kappa,0}^{1}(r_2,z)^*\chi_{\kappa}^{\mu}(\hat{\boldsymbol{r}}_2)]~.
\end{aligned}
\end{equation}

\noindent Using the definition of the multipole fields (\ref{MultComp}) and the fact that
\\
\begin{equation} \label{TensorProduct}
\begin{aligned}
\boldsymbol{\sigma}\cdot\boldsymbol{T}_{JLM} &=  \sum_\mu \langle L~(M-\mu)~1~\mu\vert J~M\rangle Y_{L,M-\mu}(\boldsymbol{\sigma}\cdot\boldsymbol{\xi}_\mu)= [\boldsymbol{Y}_L \otimes \boldsymbol{\sigma}]_{JM}~,
\end{aligned}
\end{equation}

\noindent see \cite{balashov_polarization_2000}, we can further evaluate $J_1$ and obtain for the magnetic ($P = 0$) transition

\begin{equation}
\begin{aligned}
J_1(z'\kappa'\mu',\omega,P_2 = 0,L_2M_2,z\kappa\mu, r_1) =i \Big(&K_1(z',\kappa', \omega, L_2, z, \kappa, r_1)^* \mem{\kappa'\mu'}{[\boldsymbol{Y}_{L_2} \otimes \boldsymbol{\sigma}]_{L_2M_2}}{-\kappa\mu} \\
& - K_2(z', \kappa', \omega, L_2, z, \kappa, r_1)^* \mem{-\kappa'\mu'}{[\boldsymbol{Y}_{L_2} \otimes \boldsymbol{\sigma}]_{L_2M_2}}{\kappa\mu}\Big)~,
\end{aligned}
\end{equation}

\noindent and for the electric ($P = 1$) transition

\begin{equation}
\begin{aligned}
J_1(z'\kappa'\mu',\omega,P_2 = 1,L_2M_2,z\kappa\mu, r_1) =i&\sqrt{\frac{L_2+1}{2L_2+1}} \Big(K_1(z',\kappa', \omega, L_2-1, z, \kappa, r_1)^* \mem{\kappa'\mu'}{[\boldsymbol{Y}_{L_2-1} \otimes \boldsymbol{\sigma}]_{L_2M_2}}{-\kappa\mu} \\
&\hphantom{i\sqrt{\frac{L_2+1}{2L_2+1}}} - K_2(z', \kappa', \omega, L_2-1, z, \kappa, r_1)^* \mem{-\kappa'\mu'}{[\boldsymbol{Y}_{L_2-1} \otimes \boldsymbol{\sigma}]_{L_2M_2}}{\kappa\mu}\Big)\\
&-i\sqrt{\frac{L_2}{2L_2+1}} \Big(K_1(z', \kappa', \omega, L_2+1, z, \kappa, r_1)^* \mem{\kappa'\mu'}{[\boldsymbol{Y}_{L_2+1} \otimes \boldsymbol{\sigma}]_{L_2M_2}}{-\kappa\mu} \\
&\hphantom{i\sqrt{\frac{L_2+1}{2L_2+1}}} - K_2(z', \kappa', \omega, L_2+1, z, \kappa, r_1)^* \mem{-\kappa'\mu'}{[\boldsymbol{Y}_{L_2+1} \otimes \boldsymbol{\sigma}]_{L_2M_2}}{\kappa\mu}\Big)~.
\end{aligned}
\end{equation}

\noindent Here, we have introduced the radial integrals

\begin{equation}
\begin{aligned}
K_1(z', &\kappa', \omega, L, z, \kappa, r_2) &= \int_0^{r_2} \text{d}r~r^2 F^1_{\kappa',0}(r, z') j_L (\omega r)  F^2_{\kappa,0}(r, z)~,\\
K_2(z', &\kappa', \omega, L, z, \kappa, r_2) &= \int_0^{r_2} \text{d}r~r^2 F^2_{\kappa',0}(r, z') j_L (\omega r)  F^1_{\kappa,0}(r, z)~.
\end{aligned}
\end{equation}

\noindent Now, we can insert $J_1$ back into Eq. (\ref{I1}) and by again using Eqs. (\ref{MultComp}) and (\ref{TensorProduct}), we finally obtain
\nopagebreak
\begin{equation} \label{I1Full}
\begin{aligned}
I_1(\kappa'\mu',\omega,P_1 = 0, L_1M_1,P_2 = 0,L_2M_2,\kappa\mu) = &\zeta(\omega,L_1L_1M_1,L_2L_2M_2,\kappa\mu,\kappa'\mu')~,\\
I_1(\kappa'\mu',\omega,P_1 = 1, L_1M_1,P_2 = 0,L_2M_2,\kappa\mu) = &\sqrt{\frac{L_1+1}{2L_1+1}}\zeta(\omega,L_1,L_1-1,M_1,L_2L_2M_2,\kappa\mu,\kappa'\mu')\\
&-\sqrt{\frac{L_1}{2L_1+1}}\zeta(\omega,L_1,L_1+1,M_1,L_2L_2M_2,\kappa\mu,\kappa'\mu')~,\\
I_1(\kappa'\mu',\omega,P_1 = 0, L_1M_1,P_2 = 1,L_2M_2,\kappa\mu) = &\sqrt{\frac{L_2+1}{2L_2+1}}\zeta(\omega,L_1L_1M_1,L_2,L_2-1,M_2,\kappa\mu,\kappa'\mu')\\
&-\sqrt{\frac{L_2}{2L_2+1}}\zeta(\omega,L_1L_1,M_1,L_2,L_2+1,M_2,\kappa\mu,\kappa'\mu')~,\\
I_1(\kappa'\mu',\omega,P_1 = 1, L_1M_1,P_2 = 1,L_2M_2,\kappa\mu) = &\sqrt{\frac{L_1+1}{2L_1+1}}\sqrt{\frac{L_2+1}{2L_2+1}}\zeta(\omega,L_1,L_1-1,M_1,L_2,L_2-1,M_2,\kappa\mu,\kappa'\mu')\\
&-\sqrt{\frac{L_1}{2L_1+1}}\sqrt{\frac{L_2+1}{2L_2+1}}\zeta(\omega,L_1L_1+1,M_1,L_2,L_2-1,M_2,\kappa\mu,\kappa'\mu')\\
&-\sqrt{\frac{L_1+1}{2L_1+1}}\sqrt{\frac{L_2}{2L_2+1}}\zeta(\omega,L_1L_1-1,M_1,L_2,L_2+1,M_2,\kappa\mu,\kappa'\mu')\\
&+\sqrt{\frac{L_1}{2L_1+1}}\sqrt{\frac{L_2}{2L_2+1}}\zeta(\omega,L_1L_1+1,M_1,L_2,L_2+1,M_2,\kappa\mu,\kappa'\mu')~.\\
\end{aligned}
\end{equation}

\noindent where we have introduced the function

\begin{equation}
\begin{aligned}
\zeta(&\kappa'\mu',\omega,J_1L_1M_1,J_2L_2M_2,\kappa\mu) = \\
&K_{11}(\kappa',\omega, L_1, L_2,\kappa)\mem{\kappa'\mu'}{[\boldsymbol{Y}_{L_1} \otimes \boldsymbol{\sigma}]_{J_1M_1}}{-\kappa\mu} \mem{\kappa'\mu'}{[\boldsymbol{Y}_{L_2} \otimes \boldsymbol{\sigma}]_{J_2M_2}}{-\kappa\mu}^*\\
&-K_{21}(\kappa',\omega, L_1, L_2,\kappa) \mem{-\kappa'\mu'}{[\boldsymbol{Y}_{L_1} \otimes \boldsymbol{\sigma}]_{J_1M_1}}{\kappa\mu} \mem{\kappa'\mu'}{[\boldsymbol{Y}_{L_2} \otimes \boldsymbol{\sigma}]_{J_2M_2}}{-\kappa\mu}^*\\
&-\Big[K_{12}(\kappa', \omega, L_1, L_2,\kappa) \mem{\kappa'\mu'}{[\boldsymbol{Y}_{L_1} \otimes \boldsymbol{\sigma}]_{J_1M_1}}{-\kappa\mu} \mem{-\kappa'\mu'}{[\boldsymbol{Y}_{L_2} \otimes \boldsymbol{\sigma}]_{J_2M_2}}{\kappa\mu}^*\\
&-K_{22}(\kappa',\omega, L_1, L_2,\kappa)\mem{-\kappa'\mu'}{[\boldsymbol{Y}_{L_1} \otimes \boldsymbol{\sigma}]_{J_1M_1}}{\kappa\mu} \mem{-\kappa'\mu'}{[\boldsymbol{Y}_{L_2} \otimes \boldsymbol{\sigma}]_{J_2M_2}}{\kappa\mu}^*\Big]~,
\end{aligned}
\end{equation}

\noindent as well as the radial integrals

\begin{equation} \label{KInt}
\begin{aligned}
K_{1i}(\kappa',\omega, L_1, L_2,\kappa) = &\int_{-\infty}^\infty\text{d}z'~\int_{-\infty}^\infty\text{d}z~\frac{\delta (\omega+z-z')}{w_\kappa(z)w_{\kappa'}(z')}\times\int_0^\infty\text{d}r_1~r_1^2F_{\kappa',\infty}^1(r_1,z') j_{L_1}(\omega r_1) F_{\kappa,\infty}^2(r_1,z)\\
&\times K_i(z', \kappa', \omega, L_2, z, \kappa, r_1)~,\\
K_{2i}(\kappa',\omega, L_1, L_2,\kappa) = &\int_{-\infty}^\infty\text{d}z'~\int_{-\infty}^\infty\text{d}z~\frac{\delta (\omega+z-z')}{w_\kappa(z)w_{\kappa'}(z')}\int_0^\infty\text{d}r_1~r_1^2F_{\kappa',\infty}^2(r_1,z') j_{L_1}(\omega r_1) F_{\kappa,\infty}^1(r_1,z)\\
&\times K_i(z', \kappa', \omega, L_2, z, \kappa, r_1)~. \\
\end{aligned}
\end{equation}

\noindent For $I_2$, the derivation is very similar and we must simply replace the function $\zeta$ by

\begin{equation}
\begin{aligned}
\tilde{\zeta}(&\kappa'\mu',\omega,J_1L_1M_1,J_2L_2M_2,\kappa\mu) = \\
&K_{11}(\kappa',\omega, L_2, L_1,\kappa)\mem{\kappa'\mu'}{[\boldsymbol{Y}_{L_1} \otimes \boldsymbol{\sigma}]_{J_1M_1}}{-\kappa\mu} \mem{\kappa'\mu'}{[\boldsymbol{Y}_{L_2} \otimes \boldsymbol{\sigma}]_{J_2M_2}}{-\kappa\mu}^*\\
&-K_{12}(\kappa',\omega, L_2, L_1,\kappa) \mem{-\kappa'\mu'}{[\boldsymbol{Y}_{L_1} \otimes \boldsymbol{\sigma}]_{J_1M_1}}{\kappa\mu} \mem{\kappa'\mu'}{[\boldsymbol{Y}_{L_2} \otimes \boldsymbol{\sigma}]_{J_2M_2}}{-\kappa\mu}^*\\
&-\Big[K_{21}(\kappa', \omega, L_2, L_1,\kappa) \mem{\kappa'\mu'}{[\boldsymbol{Y}_{L_1} \otimes \boldsymbol{\sigma}]_{J_1M_1}}{-\kappa\mu} \mem{-\kappa'\mu'}{[\boldsymbol{Y}_{L_2} \otimes \boldsymbol{\sigma}]_{J_2M_2}}{\kappa\mu}^*\\
&-K_{22}(\kappa',\omega, L_2, L_1,\kappa)\mem{-\kappa'\mu'}{[\boldsymbol{Y}_{L_1} \otimes \boldsymbol{\sigma}]_{J_1M_1}}{\kappa\mu} \mem{-\kappa'\mu'}{[\boldsymbol{Y}_{L_2} \otimes \boldsymbol{\sigma}]_{J_2M_2}}{\kappa\mu}^*\Big]~.
\end{aligned}
\end{equation}

\noindent in Eq. (\ref{I1Full}) to obtain the analogous expression for $I_2$. Therefore, we have reduced the problem to the calculation of the angular integrals $\mem{\kappa'\mu'}{[\boldsymbol{Y}_{L_1} \otimes \boldsymbol{\sigma}]_{J_1M_1}}{\kappa\mu}$, which are well-known and can be easily calculated analytically \cite{10.5555/1215645, Grant_1974}, and the calculation of the radial integrals $K_{ij}$ which are much more complicated. 

To further evaluate Eq. \eqref{KInt}, we have to insert the explicit form of the radial Green's function~\eqref{GreenCoulomb}. Then, we expand the four factors of the integrand, which each contain the sum of two Whittaker functions, to obtain 16 terms consisting of the product of four Whittaker functions each. For example, for $K_{11}$ we obtain

\begin{equation} \label{K11}
\begin{aligned}
K_{11}(\kappa',\omega,L_1,L_2,\kappa) &= \int_{-i\infty}^{+i\infty} \text{d}z' \frac{\sqrt{1+(z'+\frac{\omega}{2})}\sqrt{1-(z'-\frac{\omega}{2})}\sqrt{1+(z'+\frac{\omega}{2})}\sqrt{1-(z'-\frac{\omega}{2})}}{16c^2c'^2}\\
&\times [\mathcal{I}(z',\kappa',\omega,L_1,L_2,\kappa,-1,-1,-1,-1)-\mathcal{I}(z',\kappa',\omega,L_1,L_2,\kappa,-1,-1,+1,-1)\\
&+\mathcal{I}(z',\kappa',\omega,L_1,L_2,\kappa,-1,-1,-1,+1)-\mathcal{I}(z',\kappa',\omega,L_1,L_2,\kappa,-1,-1,+1,+1)\\
&+\mathcal{I}(z',\kappa',\omega,L_1,L_2,\kappa,+1,-1,-1,-1)-\mathcal{I}(z',\kappa',\omega,L_1,L_2,\kappa,+1,-1,+1,-1)\\
&+\mathcal{I}(z',\kappa',\omega,L_1,L_2,\kappa,+1,-1,-1,+1)-\mathcal{I}(z',\kappa',\omega,L_1,L_2,\kappa,+1,-1,+1,+1)\\
&-(\mathcal{I}(z',\kappa',\omega,L_1,L_2,\kappa,-1,+1,-1,-1)-\mathcal{I}(z',\kappa',\omega,L_1,L_2,\kappa,-1,+1,+1,-1)\\
&+\mathcal{I}(z',\kappa',\omega,L_1,L_2,\kappa,-1,+1,-1,+1)-\mathcal{I}(z',\kappa',\omega,L_1,L_2,\kappa,-1,+1,+1,+1))\\
&-(\mathcal{I}(z',\kappa',\omega,L_1,L_2,\kappa,+1,+1,-1,-1)-\mathcal{I}(z',\kappa',\omega,L_1,L_2,\kappa,+1,+1,+1,-1)\\
&+\mathcal{I}(z',\kappa',\omega,L_1,L_2,\kappa,+1,+1,-1,+1)-\mathcal{I}(z',\kappa',\omega,L_1,L_2,\kappa,+1,+1,+1,+1))]~,
\end{aligned}
\end{equation}

\noindent where

\begin{equation}
\begin{aligned}
\mathcal{I}(&z',\kappa',\omega,L_1,L_2,\kappa, p_1, p_2, p_3, p_4) = C_{p_1}C_{p_2}' \tilde{C}_{p_3}\tilde{C}_{p_4}'\mathcal{J}(z',\kappa',\omega,L_1,L_2,\kappa,p_1,p_2,p_3,p_4)~,
\end{aligned}
\end{equation}

\noindent and

\begin{equation}
\begin{aligned}
C_{-1} &= \kappa+\frac{\gamma}{c},~C_{+1} = 1~,~C_{-1}' = \kappa'+\frac{\gamma}{c'},~C_{+1}' = 1~,\\
\tilde{C}_{-1} &= \lambda-\nu,~C_{+1} = \kappa-\frac{\gamma}{c}~,~\tilde{C}_{-1}' = \lambda'-\nu',~C_{+1}' = \kappa'-\frac{\gamma}{c'}~,\\
\end{aligned}
\end{equation}

\noindent as well as

\begin{equation}
\begin{aligned}
\mathcal{J}(z',\kappa',\omega,L_1,L_2,\kappa,p_1,p_2,p_3,p_4) = &\frac{\Gamma(\lambda-\nu)\Gamma(\lambda'-\nu')}{\Gamma(1+2\lambda)\Gamma(1+2\lambda')}\int_0^\infty \frac{\text{d}r_1}{r_1} W_{\nu'+\frac{p_1}{2},\lambda'}(2c'r_1) j_{L_1}(\omega r_1) W_{\nu+\frac{p_2}{2},\lambda}(2cr_1)\\
&\times\int_0^{r_1} \frac{\text{d}r_2}{r_2}M_{\nu'+\frac{p_3}{2},\lambda'}(2c'r_1) j_{L_2}(\omega r_1) M_{\nu+\frac{p_4}{2},\lambda}(2cr_1)~.\\
\end{aligned}
\end{equation}

\noindent Here, in Eq.~\eqref{K11}, we have already performed the integration over $z$ as discussed in Sec. \ref{EInt}. We can write the other integrals $K_{12}$, $K_{21}$ and $K_{22}$ very similarly where the $K_{ij}$ only differ by the signs between the individual terms in Eq.~(\ref{K11}) and the prefactor. Therefore, the problem is finally reduced to the evaluation of the integral $\mathcal{J}(\kappa',\omega,L_1,L_2,\kappa,z',p_1,p_2,p_3,p_4)$ for all 16 possible combinations of $p_1, p_2, p_3, p_4 = \pm 1$.

\section{Analytical solution of the radial integrals}
In this section, we derive the analytical solution of the radial integrals in Eq.~\eqref{IntSplit}. We first insert the asymptotic expansion of the Whittaker functions~\eqref{asympWhit} and of the spherical Bessel function~\eqref{asympBes} into the third and fourth line of Eq.~\eqref{IntSplit}. We omit the second term in the asymptotic expansion of $M_{a,b}(z)$ since it smaller by a factor $e^z$ which is around 500 orders of magnitude for typical values of $a$ and $a'$ used in our calculation. However, since this term has the same form as the first one, including it is completely analogous to the following derivation. We obtain for the integral

\begin{equation} \label{asympInt}
\begin{aligned}
&\int_a^\infty \frac{\text{d}r_1}{r_1} W_{\nu'+\frac{p_1}{2},\lambda'}(2c'r_1) j_{L_1}(\omega r_1) W_{\nu+\frac{p_2}{2},\lambda}(2cr_1)\Big[\mathcal{C} +\int_{a'}^{r_1} \frac{\text{d}r_2}{r_2} M_{\nu'+\frac{p_3}{2},\lambda'}(2c'r_2) j_{L_2}(\omega r_2) M_{\nu+\frac{p_4}{2},\lambda}(2cr_2)\Big]\\
&\to \sum_{s_W', s_W = 0}^\infty u_W(s_W', \nu'+\frac{p_1}{2}, \lambda')u_W(s_W, \nu+\frac{p_2}{2}, \lambda)\int_a^\infty \frac{\text{d}r_1}{r_1} e^{-(c+c')r_1} (2c'r_1)^{\nu'+p_1/2}j_{L_1}(\omega r_1)(2cr_1)^{\nu+p_2/2}\\
&\times (-2c'r_1)^{-s_W'}(-2cr_1)^{-s_W}\Big[\mathcal{C} + \frac{\Gamma(1+2\lambda)\Gamma(1+2\lambda')}{\Gamma(\frac{1}{2}+\lambda-\nu-\frac{p_4}{2})\Gamma(\frac{1}{2}+\lambda'-\nu'-\frac{p_3}{2})}\sum_{s_M', s_M = 0}^\infty u_M(s_M',\nu'+\frac{p_3}{2},\lambda')\\
&\times u_M(s_M,\nu+\frac{p_4}{2},\lambda)\int_{a'}^{r_1} \frac{\text{d}r_2}{r_2} e^{+(c+c')r_2} (2c'r_2)^{-\nu'-p_3/2}j_{L_2}(\omega r_2)(2cr_2)^{-\nu-p_4/2}(2c'r_2)^{-s_M'}(2cr_2)^{-s_M}\Big]\\
&= (2c')^{\nu'+p_1/2}(2c)^{\nu+p_2/2}\bigg\{\text{I}\times\Bigg[\mathcal{C} - \frac{\Gamma(1+2\lambda)\Gamma(1+2\lambda')(2c')^{-\nu'-p_3/2}(2c)^{-\nu-p_4/2}}{\Gamma(\frac{1}{2}+\lambda-\nu-\frac{p_4}{2})\Gamma(\frac{1}{2}+\lambda'-\nu'-\frac{p_3}{2})}\times \text{II}\Bigg]\\
&+\frac{\Gamma(1+2\lambda)\Gamma(1+2\lambda')(2c')^{-\nu'-p_3/2}(2c)^{-\nu-p_4/2}}{\Gamma(\frac{1}{2}+\lambda-\nu-\frac{p_4}{2})\Gamma(\frac{1}{2}+\lambda'-\nu'-\frac{p_3}{2})}\times\text{III} \bigg\}~,
\end{aligned}
\end{equation}

\noindent where

\begin{equation} \label{IntToSolve}
\begin{aligned}
\text{I} = &\sum_{s_W', s_W = 0}^\infty (-2c')^{-s_W'} u_W(s_W', \nu'+\frac{p_1}{2}, \lambda')(-2c)^{-s_W}u_W(s_W, \nu+\frac{p_2}{2}, \lambda)\\
&\times\int_a^\infty \text{d}r_1~e^{-(c+c')r_1} r_1^{-1+\nu'+\nu+(p_1+p_2)/2-s_W'-s_W}j_{L_1}(\omega r_1)~,\\
\text{II} = &\sum_{s_M', s_M = 0}^\infty (2c')^{-s_M'}u_M(s_M',\nu'+\frac{p_3}{2},\lambda')(2c)^{-s_M}u_M(s_M,\nu+\frac{p_4}{2},\lambda)\\
&\times\Bigg[\left.\int \text{d}r_2~e^{+(c+c')r_2} r_2^{-1-\nu'-\nu-(p_3+p_4)/2-s_M'-s_M}j_{L_2}(\omega r_2)\Bigg]\right\vert_{r_2=a'}~,\\
\text{III} = &\sum_{s_W', s_W, s_M', s_M = 0}^\infty(-2c')^{-s_W'} u_W(s_W', \nu'+\frac{p_1}{2}, \lambda')(-2c)^{-s_W}u_W(s_W, \nu+\frac{p_2}{2}, \lambda)(2c')^{-s_M'}u_M(s_M',\nu'+\frac{p_3}{2},\lambda') \\
&\times(2c)^{-s_M}u_M(s_M,\nu+\frac{p_4}{2},\lambda)\int_a^\infty \text{d}r_1~e^{-(c+c')r_1} r_1^{-1+\nu'+\nu+(p_1+p_2)/2-s_W'-s_W}j_{L_1}(\omega r_1)\\
&\times \Bigg[\left.\int \text{d}r_2~e^{+(c+c')r_2} r_2^{-1-\nu'-\nu-(p_3+p_4)/2-s_M'-s_M}j_{L_2}(\omega r_2)\Bigg]\right\vert_{r_2=r_1}~,
\end{aligned}
\end{equation}

\noindent and

\begin{equation}
\mathcal{C} = \int_0^{a'} \frac{\text{d}r_2}{r_2} M_{\nu'+\frac{p_3}{2},\lambda'}(2c'r_2) j_{L_2}(\omega r_2) M_{\nu+\frac{p_4}{2},\lambda}(2cr_2)~.
\end{equation}

\noindent \textbf{Analytical solution for $\omega \neq 0$}\\
\noindent To evaluate these integrals for $\omega \neq 0$, we use the exact expansion of the spherical Bessel function~\eqref{asympBes} and obtain for the integral over $r_2$

\begin{equation} \label{r1Int}
\begin{aligned}
\int&\text{d}r_2~e^{+(c+c')r_2} r_2^{-1-\nu'-\nu-(p_3+p_4)/2-s_M'-s_M}j_{L_2}(\omega r_2) \\
&= \sum_{m_2=0}^{L_2} \frac{(L_2+m_2)!}{m_2!(L_2-m_2)!} i^{L_2+1-m_2} (2\omega)^{-m_2-1}[(-1)^{L_2+1-m_2}N_++N_-]~,
\end{aligned}
\end{equation}

\noindent where

\begin{equation}\label{Npm}
\begin{aligned}
N_{\pm} &= \int\text{d}r_2~e^{+(c+c'\pm i\omega)r_2} r_2^{-2-\nu'-\nu-p_3/2-p_4/2-m_2-s_M'-s_M} \\
&= \frac{(-c-c'\mp i\omega)^{2+\nu+\nu'+(p_3+p_4)/2+m_2+s_M'+s_M}}{c+c'\pm i\omega}\\
&\times \Gamma(-1-\nu'-\nu-(p_3+p_4)/2-m_2-s_M'-s_M, -(c+c'\pm i\omega)r_2)\\
&\to\sum_{s_G = 0}^\infty \frac{(2+\nu'+\nu+p_3/2+p_4/2+m_2+s_M'+s_M)_{s_G}}{(c+c'\pm i\omega)^{s_G+1}} r_2^{-2-\nu'-\nu-(p_3+p_4)/2-m_2-s_M'-s_M-s_G}e^{(c+c'\pm i\omega)r_2}~.
\end{aligned}
\end{equation}

\noindent Here, in the last step, we have replaced the incomplete gamma function by its full asymptotic expansion

\begin{equation}
\Gamma(a,z) = z^{a-1} e^{-z} \sum_{s=0}^\infty \frac{u_G(s,a)}{z^s} ~,
\end{equation}

\noindent where

\begin{equation}
u_G(s,a) = (-1)^s(1-a)_s~.
\end{equation}

If we set $r_2 = a'$ in Eq.~\eqref{Npm}, we can easily calculate II in Eq. \eqref{IntToSolve}. To evaluate III, we have to set $r_2 = r_1$ and integrate over $r_1$

\begin{equation}
\begin{aligned}
&\int_a^\infty \text{d}r_1~e^{-(c+c')r_1} r_1^{-1+\nu'+\nu+(p_1+p_2)/2-s_W'-s_W}j_{L_1}(\omega r_1) \Bigg[\left.\int \text{d}r_2~e^{+(c+c')r_2} r_2^{-1-\nu'-\nu-(p_3+p_4)/2-s_M'-s_M}j_{L_2}(\omega r_2)\Bigg]\right\vert_{r_2=r_1}\\
&= \sum_{m_1=0}^{L_1} \sum_{m_2=0}^{L_2} \frac{(L_1+m_1)!}{m_1!(L_1-m_1)!} \frac{(L_2+m_2)!}{m_2!(L_2-m_2)!} i^{L_1+L_2+2-m_1-m_2} (2\omega)^{-m_1-m_2-2}\\
&\times [(-1)^{L_1+L_2+2-m_1-m_2}N_{++}+(-1)^{L_1+1-m_1}N_{+-}+(-1)^{L_2+1-m_2}N_{-+}+N_{--}]~,
\end{aligned}
\end{equation}

\noindent where

\begin{align}
N_{++} &= \sum_{s_G=0}^\infty \int_a^\infty \text{d}r_1~\frac{(2+\nu'+\nu+(p_3+p_4)/2+m_2+s_M'+s_M)_{s_G}}{(c+c'+ i\omega)^{s_G+1}}\nonumber\\
&\times e^{2i\omega r_1}r_1^{-4+(p_1+p_2-p_3-p_4)/2-m_1-m_2-s_M'-s_M-s_W'-s_W-s_G}\nonumber \\
&= \sum_{s_G=0}^\infty-\frac{(2+\nu'+\nu+(p_3+p_4)/2+m_2+s_M'+s_M)_{s_G}}{2i\omega(c+c'+ i\omega)^{s_G+1}}\nonumber\\
&\times (-2i\omega)^{4-(p_1+p_2-p_3-p_4)/2+m_1+m_2+s_M'+s_M+s_W'+s_W+s_G}\nonumber\\
&\times\Gamma(-3+(p_1+p_2-p_3-p_4)/2-m_1-m_2-s_M'-s_M-s_W'-s_W-s_G,-2i\omega a)~,\nonumber\\
N_{--} &= \sum_{s_G=0}^\infty \int_a^\infty\text{d}r_1~\frac{(2+\nu'+\nu+(p_3+p_4)/2+m_2+s_M'+s_M)_{s_G}}{(c+c'- i\omega)^{s_G+1}}\nonumber \\
&\times e^{-2i\omega r_1}r_1^{-4+(p_1+p_2-p_3-p_4)/2-m_1-m_2-s_M'-s_M-s_W'-s_W-s_G}\nonumber\\
&= \sum_{s_G=0}^\infty+\frac{(2+\nu'+\nu+(p_3+p_4)/2+m_2+s_M'+s_M)_{s_G}}{2i\omega(c+c'- i\omega)^{s_G+1}}\nonumber\\
&\times(2i\omega)^{4-(p_1+p_2-p_3-p_4)/2+m_1+m_2+s_M'+s_M+s_W'+s_W+s_G}\nonumber\\
&\times \Gamma(-3+(p_1+p_2-p_3-p_4)/2-m_1-m_2-s_M'-s_M-s_W'-s_W-s_G,+2i\omega a)~,\nonumber\\
N_{+-} &= \sum_{s_G=0}^\infty\int_a^\infty\text{d}r_1~\frac{(2+\nu'+\nu+(p_3+p_4)/2+m_2+s_M'+s_M)_{s_G}}{(c+c'- i\omega)^{s_G+1}} r_1^{-4+(p_1+p_2-p_3-p_4)/2-m_1-m_2-s_M'-s_M-s_W'-s_W-s_G}\nonumber\\
& = \sum_{s_G=0}^\infty \frac{(2+\nu'+\nu+(p_3+p_4)/2+m_2+s_M'+s_M)_{s_G}a^{-3+(p_1+p_2-p_3-p_4)/2-m_1-m_2-s_M'-s_M-s_W'-s_W-s_G}}{(c+c'- i\omega)^{s_G+1}(3-(p_1+p_2-p_3-p_4)/2+m_1+m_2+s_M'+s_M+s_W'+s_W+s_G)}~, \nonumber\\
N_{-+} &= \sum_{s_G=0}^\infty\int_a^\infty\text{d}r_1~\frac{(2+\nu'+\nu+(p_3+p_4)/2+m_2+s_M'+s_M)_{s_G}}{(c+c'+ i\omega)^{s_G+1}} r_1^{-4+(p_1+p_2-p_3-p_4)/2-m_1-m_2-s_M'-s_M-s_W'-s_W-s_G}\nonumber\\
& = \sum_{s_G=0}^\infty \frac{(2+\nu'+\nu+(p_3+p_4)/2+m_2+s_M'+s_M)_{s_G}a^{-3+(p_1+p_2-p_3-p_4)/2-m_1-m_2-s_M'-s_M-s_W'-s_W-s_G}}{(c+c'+ i\omega)^{s_G+1}(3-(p_1+p_2-p_3-p_4)/2+m_1+m_2+s_M'+s_M+s_W'+s_W+s_G)}~. \nonumber\\
\end{align}

Finally, we obtain for the integral over $r_2$ in I in Eq. \eqref{IntToSolve}

\begin{equation}
\begin{aligned}
\int_a^\infty&\text{d}r_1~e^{-(c+c')r_1} r_1^{-1+\nu'+\nu+(p_1+p_2)/2-s_W'-s_W}j_{L_1}(\omega r_1) \\
&= \sum_{m_1=0}^{L_1} \frac{(L_1+m_1)!}{m_1!(L_1-m_1)!} i^{L_1+1-m_1} (2\omega)^{-m_1-1}[(-1)^{L_1+1-m_1}O_++O_-]~,
\end{aligned}
\end{equation}

\noindent where

\begin{equation}
\begin{aligned}
O_{\pm} &= \int_a^\infty\text{d}r_1~e^{-(c+c'\mp i\omega)r_1} r_1^{-2+\nu'+\nu+(p_1+p_2)/2-m_1-s_W-s_W'} \\
&= \frac{(c+c'\mp i\omega)^{2-\nu-\nu'-(p_1+p_2)/2+m_1+s_W'+s_W}}{c+c'\mp i\omega}\\
&\times \Gamma(-1+\nu'+\nu+(p_1+p_2)/2-m_1-s_W'-s_W, +(c+c'\mp i\omega)a)\\
&\to\sum_{s_G = 0}^\infty (-1)^{s_G} \frac{(2-\nu'-\nu-(p_1+p_2)/2+m_1+s_W'+s_W)_{s_G}}{(c+c'\mp i\omega)^{s_G+1}}a^{-2+\nu'+\nu+(p_1+p_2)/2-m_1-s_W'-s_W-s_G}e^{-(c+c'\mp i\omega)a}~.
\end{aligned}
\end{equation}

It can be easily seen from the equations above, that the only term that results in a logarithmically divergent energy integral is the one for $p1,p2 = +1, p3,p4 = -1$. Since this term does not dependent on $Z$, it is trivial to show that the subtraction of the free loop diagram cancels the divergence.\\

\noindent \textbf{Analytical solution for $\omega = 0$}

\noindent For $\omega = 0$, the spherical Bessel function reduces to $j_L(0) = \delta_{L,0}$ and, hence, the integration is much easier. We obtain for the integral over $r_2$

\begin{equation} \label{r1Intw0}
\begin{aligned}
\int&\text{d}r_2~e^{+(c+c')r_2} r_2^{-1-\nu'-\nu-(p_3+p_4)/2-s_M'-s_M}j_{L_2}(0) \\
&= \delta_{L_2,0} \frac{(-c-c')^{1+\nu+\nu'+(p_3+p_4)/2+s_M'+s_M}}{c+c'}\Gamma(-\nu'-\nu-(p_3+p_4)/2-s_M'-s_M, -(c+c')r_2)\\
&\to \delta_{L_2,0} \sum_{s_G = 0}^\infty \frac{(1+\nu'+\nu+p_3/2+p_4/2+s_M'+s_M)_{s_G}}{(c+c')^{s_G+1}} r_2^{-1-\nu'-\nu-(p_3+p_4)/2-s_M'-s_M-s_G}e^{(c+c')r_2}~.
\end{aligned}
\end{equation}

\noindent Here, again, in the last step, we have replaced the incomplete gamma function by its full asymptotic expansion. As before, if we set $r_2 = a'$ in Eq.~\eqref{r1Intw0}, we can easily calculate II in Eq. \eqref{IntToSolve}. To evaluate III, we have to set $r_2 = r_1$ and integrate over $r_1$

\begin{equation}
\begin{aligned}
&\int_a^\infty \text{d}r_1~e^{-(c+c')r_1} r_1^{-1+\nu'+\nu+(p_1+p_2)/2-s_W'-s_W}j_{L_1}(0) \Bigg[\left.\int \text{d}r_2~e^{+(c+c')r_2} r_2^{-1-\nu'-\nu-(p_3+p_4)/2-s_M'-s_M}j_{L_2}(0)\Bigg]\right\vert_{r_2=r_1}\\
&= \delta_{L_1,0}\delta_{L_2,0}\sum_{s_G = 0}^\infty \frac{(1+\nu'+\nu+p_3/2+p_4/2+s_M'+s_M)_{s_G}}{(c+c')^{s_G+1}} \int_a^\infty \text{d}r_1~ r_1^{-2+(p_1+p_2-p_3-p_4)/2-s_W'-s_W-s_M'-s_M-s_G}\\
&= -\delta_{L_1,0}\delta_{L_2,0}\sum_{s_G = 0}^\infty \frac{(1+\nu'+\nu+p_3/2+p_4/2+s_M'+s_M)_{s_G}}{(c+c')^{s_G+1}}\\
&\times \frac{a^{-1+(p_1+p_2-p_3-p_4)/2-s_W'-s_W-s_M'-s_M-s_G}}{-1+(p_1+p_2-p_3-p_4)/2-s_W'-s_W-s_M'-s_M-s_G}~.
\end{aligned}
\end{equation}

Finally, we obtain for the integral over $r_1$ in I in Eq. \eqref{IntToSolve}

\begin{equation}
\begin{aligned}
\int_a^\infty&\text{d}r_1~e^{-(c+c')r_1} r_1^{-1+\nu'+\nu+(p_1+p_2)/2-s_W'-s_W}j_{L_1}(0) \\
&= \delta_{L_1,0}\frac{(c+c')^{1-\nu-\nu'-(p_1+p_2)/2+s_W'+s_W}}{c+c'} \Gamma(\nu'+\nu+(p_1+p_2)/2-s_W'-s_W, +(c+c')a)\\
&\to\delta_{L_1,0}\sum_{s_G = 0}^\infty (-1)^{s_G} \frac{(1-\nu'-\nu-(p_1+p_2)/2+s_W'+s_W)_{s_G}}{(c+c')^{s_G+1}}a^{-1+\nu'+\nu+(p_1+p_2)/2-s_W'-s_W-s_G}e^{-(c+c')a}~.
\end{aligned}
\end{equation}
\end{widetext}

\end{document}